\documentclass[
  reprint,
  aps,
  amsmath,
  amssymb,
  superscriptaddress,
  prl,
  10pt,
]{revtex4-2}

\usepackage{graphicx} 
\usepackage{dcolumn}  
\usepackage{bm}       
\begin{document}

\title{Efficient laser ion acceleration in near-critical density plasmas in the picosecond pulse regime}

\author{J. Luoma}
\affiliation{School of Applied \& Engineering Physics, Cornell University}
\author{A. Kemp}
\affiliation{Lawrence Livermore National Laboratory}
\author{A. Longman}
\affiliation{Lawrence Livermore National Laboratory}
\author{D. Rusby}
\affiliation{Lawrence Livermore National Laboratory}
\author{G. Shvets}
\affiliation{School of Applied \& Engineering Physics, Cornell University}

\begin{abstract}
Kilojoule-class short-pulse ($<10$~ps) lasers are excellent tools for generating relativistic ion beams, but achieving high laser-to-ion conversion efficiency remains an open challenge. Recent experiments show a significant increase in conversion efficiency when leveraging targets with an average plasma density approaching the relativistic-critical limit. We present an analytical model that explains this increase by extending the theory of target-normal sheath acceleration (TNSA) to include laser penetration of relativistic-critical density targets, enabling energy transfer to the ion front by heating the expanding electron sheath. This physical process increases the sheath field strength and significantly improves ion cutoff energy and laser-to-ion conversion efficiency relative to classical TNSA. The model predicts a maximum ion conversion efficiency of $37\%$ by optimizing laser transmission and target areal density. Key scalings of the model agree with particle-in-cell (PIC) simulations and published experimental data. A series of 2D and 3D simulations of hydrocarbon plasmas confirm the robustness of the acceleration process, demonstrating a path for maximizing ion yields using relativistic-critical targets.
\end{abstract}

\maketitle

High-power lasers provide a unique capacity to drive high-current pulses of MeV-energy ions \cite{snavely2000intense} that are well suited for applications including isochoric heating \cite{patel2003isochoric,bailly2025creation}, radiography \cite{li2006measuring,mackinnon2006proton}, and cancer therapy \cite{bulanov2002oncological,kroll2022tumour,bin2022new,flacco2025laser}. While many of these applications benefit from high ion conversion efficiency, laser-driven acceleration is often viewed as an inefficient process due to the slow response of ions in optical-frequency laser fields.

Previous studies have proposed a variety of ion acceleration mechanisms to deliver large fluxes of high-energy ions. Target-normal sheath acceleration (TNSA) is a robust technique, but conversion efficiencies in foil targets are generally limited to 1-3\% with few measurements beyond 5\% \cite{fuchs2006laser,robson2007scaling,tochitsky2025high,snavely2000intense,wagner2016maximum}. Breakout-afterburner (BOA) and radiation-pressure acceleration (RPA) provide routes to higher conversion efficiency; however, these mechanisms typically require thin foils (10's to 100's nm) and demand high laser contrast to preserve the target prior to the main laser drive \cite{henig2009enhanced,yin2011three,yin2011break,powell2015proton}. Such constraints are difficult to achieve on the kJ-class laser systems required to drive high ion flux.

Acceleration in the relativistic-transparency regime offers high coupling efficiency as demonstrated in experiments and simulations \cite{willingale2009characterization,willingale2011high,higginson2018near,rehwald2023ultra}. Explanations for the enhanced efficiency range from TNSA enhancement to a mix of RPA, TNSA, and BOA \cite{powell2015proton,gonzalez2016towards}. While relativistic-transparency is a promising route for efficient laser-driven ion acceleration, a concise framework is necessary to understand the scaling and optimization of this mechanism.

\begin{figure*}[t]
    \centering
    \includegraphics[width=\textwidth]{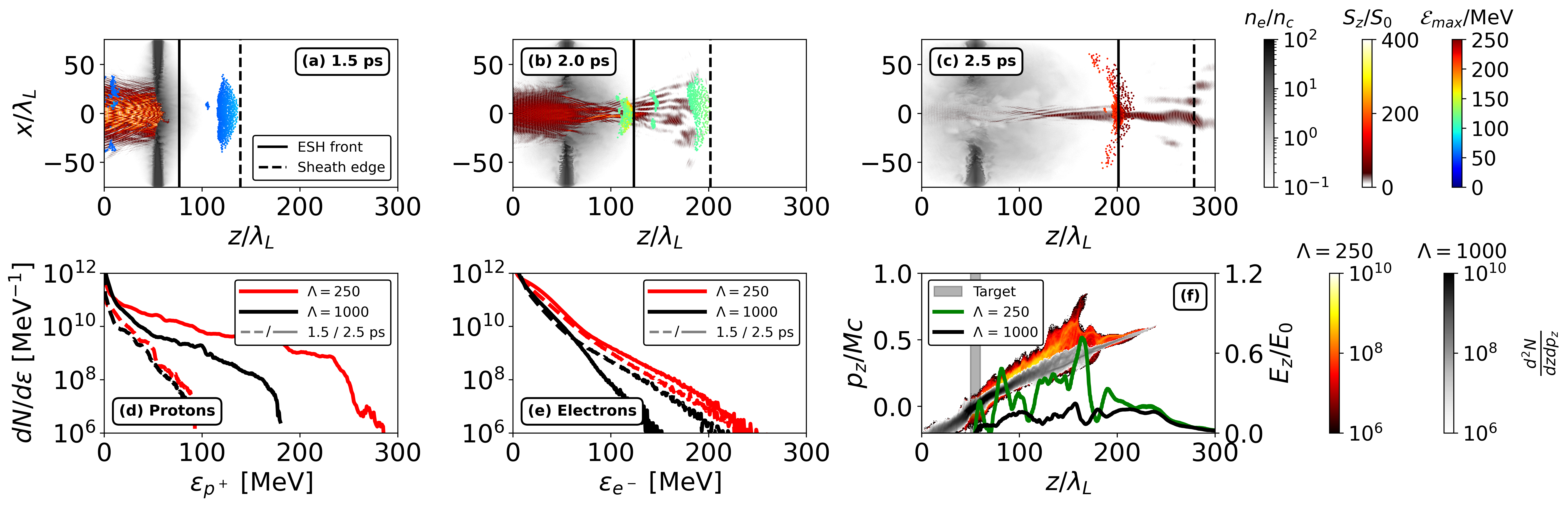}
    \caption{Temporal evolution of a 5.5-kJ (200-J/\textmu m in 2D) laser interacting with an ESH-optimized target ($n_{e0}/n_c=25$, $L_0/\lambda_L=10$, $\Lambda=250$). Panels (a)–(c) show electron density (gray), Poynting flux (orange), and the position and energy of the highest-energy protons (blue-red points) for $t=1.5$ to 2.5~ps. Solid and dashed vertical lines mark the ESH front and sheath edge. Panels (d) and (e) provide proton and electron spectra for an optimized ($\Lambda=250$) and opaque ($\Lambda=1000$) target. Panel (f) shows proton longitudinal phase space and on-axis sheath field at $t=2.25$~ps.
    }
    \label{2D_PIC}
\end{figure*}

We present a 1D analytical model for a robust ion acceleration mechanism driven by transmitting laser energy into an expanding electron sheath, hereafter referred to as extended sheath heating (ESH). A series of particle-in-cell (PIC) simulations verify the model and its key predictions. The ESH process begins with the expansion of a hot plasma. Fast electron currents establish a strong sheath that launches and accelerates ions; however, the hot electron population will adiabatically cool in the absence of a laser drive \cite{mora2009rarefaction,diaw2012thin}. Transmission of laser energy through the plasma bulk enables continuous heating of sheath electrons which, in turn, accelerates ions with greater efficiency \cite{mishra2018enhanced}. The essential physics of ESH is captured analytically by coupling transmission-based electron heating to a self-similar plasma expansion. 

Assuming linear absorption within a uniform, step-like plasma of initial electron density $n_{e0}$ and length $L_{0}$, the laser intensity $I_{L}$ satisfies \cite{rybicki1979radiative}

\begin{equation}
    \frac{dI_{L}}{dz}=-\alpha I_{L}.
    \label{attenuation_ODE}
\end{equation}

Here $\alpha$ is an attenuation coefficient and, as derived in Supplemental Material \cite{supplemental} (see Refs. \cite{iwata2025energy,rajawat2016one,stark2018detailed,nakamura2010high,tazes2024efficient} therein), can be expressed in terms of initial conditions for the target and laser. The transmitted laser heats electrons in the rear plasma sheath. Since acceleration of electrons and ions is linearly dependent on field strength, rather than intensity, we introduce a field transmission coefficient defined as $T=\sqrt{I_{L}/I_{L0}}$ where $I_{L0}$ is the initial laser intensity with a corresponding normalized vector potential $a_0$. Thus, the transmitted field is $Ta_0$ where

\begin{equation}
    T = \exp\left(-\frac{\alpha L_0}{2}\right)=\exp\left(-\frac{\Lambda}{C_{d}\Theta_{0}}\right).
    \label{attenuation}
\end{equation}

We define $\Theta_{0}$ as the electron temperature, normalized to $m_ec^2$, corresponding to initial laser conditions and $\Lambda=n_{e0}L_{0}/n_{c}\lambda_{L}$ as a normalized areal density accounting for target conditions. Parameters $\lambda_{L}$ and $n_{c}=4\pi^2\epsilon_{0}m_{e}c^2/q_{e}^2\lambda_{L}^2$ are laser wavelength and the critical plasma density, respectively. Constants $\epsilon_{0}$, $m_{e}$, $c$, and $q_{e}$ are vacuum permittivity, electron mass, speed of light, and electron charge.  The constant $C_{d}$ accounts for integration over the laser spatiotemporal profile. In this work, $C_{2D}=\pi/\sqrt{2\ln{2}}$ is used for 2D Gaussian envelopes.

Provided a laser field $a_0>1$ and an intensity full width at half maximum (FWHM) $\tau_{p}\gg\lambda_{L}/c$, heating of the electron sheath is well described by the stochastic acceleration model introduced by Miller et al. \cite{miller2023maximizing} as discussed in Supplemental Material \cite{supplemental}. The normalized temperature $\Theta_{e}$ of the sheath scales with transmitted field $Ta_0$,

\begin{equation}
    \Theta_{e}=\frac{Ta_{0}}{2}\sqrt{\frac{c\tau_{p}}{\pi\lambda_{L}}}.
    \label{sheath_temp}
\end{equation}

In the limit of low transmission where $\Theta_{e}\le~\Theta_{pond}=\sqrt{1+a_{0}^{2}/2}-1$, we default to the ponderomotive temperature $\Theta_{e}=\Theta_{pond}$. Similarly, $\Theta_{0}=\Theta_{e}\left(T=1\right)$ is the electron temperature under the action of the initial laser field $a_0$, as appears in Eq.~\ref{attenuation}.

Assuming electrons follow a Maxwell-Boltzmann distribution and an electrostatic sheath, the electron density profile is given by

\begin{equation}
    n_{e} = n_{e0}\exp\left(\frac{|q_e|\Phi}{\Theta_{e}m_{e}c^2}\right).
    \label{Boltzmann}
\end{equation}

Differentiating the potential $\Phi$ gives the sheath field $E_z$ \cite{mora2003plasma},

\begin{equation}
    |q_e|E_{z} =-|q_e|\partial_{z}\Phi = -\Theta_{e}m_{e}c^{2}\partial_{z}\ln{\left(n_e/n_{e0}\right)}.
    \label{sheath_eqn}
\end{equation}

Eq.~\ref{sheath_eqn} shows the transmitted laser enhances the sheath field by $\Theta_e=T\Theta_0$, which can greatly exceed the ponderomotive temperature. The impact of transmission on ion acceleration is modeled by coupling $E_z$ to the ion momentum equation,
\vspace{2pt}
\begin{equation}
    \partial_t v_i + v_i\partial_z v_i = \frac{Z|q_e|}{M}E_z.
    \label{mom_eqn}
\end{equation}

Ions are modeled as a single species of charge $Z$, mass $M$, and velocity $v_i$. The ion momentum equation has established self-similar solutions given by $v_i=z/t+c_s$ and $\ln{(n_{e}/n_{e0})}=-z/c_st-1$, where $c_s=\sqrt{Z\Theta_{e}m_{e}c^{2}/M}$ is the sound speed \cite{mora2003plasma}. While previous publications have explored these solutions in the context of TNSA, coupling the plasma expansion to transmission-based electron heating is unique to this work and critical for capturing the ESH effect \cite{gurevich1981ion,mora2003plasma,iwata2017fast}.

Enforcing quasi-neutrality via $n_e(z)=Zn_i(z)$, the self-similar solutions yield an energy spectrum for ions after an acceleration period $t_{acc}$. Note the spectrum is in units of ions per unit area,

\begin{equation}
    \frac{dN}{d\mathcal{E}_{i}} = \frac{n_{i0}t_{acc}}{\sqrt{2\mathcal{E}_{i}M}}\exp\left(-\sqrt{\frac{2\mathcal{E}_i}{Z\Theta_{e}m_{e}c^2}}\right).
    \label{ion_energy_spectrum}
\end{equation}

We consider mass-limited targets by imposing the constraint $N_{i} \le n_{i0}L_{0}$ where $N_{i}$ is the total number of ions per unit area. This constraint is satisfied when $c_{s}t_{acc} \le L_{0}$. The laser-to-ion conversion efficiency $\eta$ for the ESH process is obtained by integrating the energy-weighted ion spectrum $\mathcal{E}_{i}\frac{dN}{d\mathcal{E}_{i}}$ and normalizing to the laser energy,

\begin{equation}
    \eta = \frac{c_{s}t_{acc}}{L_{0}}T\ln{\left(T^{-1} \right)}.
    \label{ion_efficiency}
\end{equation}

\vspace{14pt}
Notably in the mass-limited regime where $c_{s}t_{acc}=L_{0}$, $\eta$ is simply a function of $T$,

\begin{equation}
    \eta=T\ln{\left(T^{-1} \right)}.
    \label{ion_efficiency_ml}
\end{equation}

Under this approximation, $\eta$ has a well-defined maximum at $T^{*}=e^{-1}$ yielding a peak conversion efficiency of $\eta^{*}\approx~37\%$. The asterisk $^{*}$ denotes values that maximize $\eta$, corresponding to optimal conditions for ion acceleration with the ESH process. The transmission scaling and optimization of $\eta(T)$ is a key finding of this work and can be verified with PIC simulations and experimental data.

The transmission coefficient $T$ has a unique one-to-one mapping to $\Lambda$ through the attenuation law given by Eq.~\ref{attenuation}. Hence, $\eta$ is equivalently maximized by an optimal areal density $\Lambda^{*}=C_{d}\Theta_{0}$. Substituting the stochastic heating model for $\Theta_{0}$, we arrive at an expression for the optimum areal density based solely on initial laser conditions,

\begin{equation}
    \Lambda^{*} = \left(\frac{n_eL_0}{n_c\lambda_{L}}\right)_{opt} = C_{d}\frac{a_{0}}{2}\sqrt{\frac{c\tau_{p}}{\pi\lambda_{L}}}.
    \label{optimal_AD}
\end{equation}

\begin{figure}[t]
    \centering
    \includegraphics[scale=0.55]{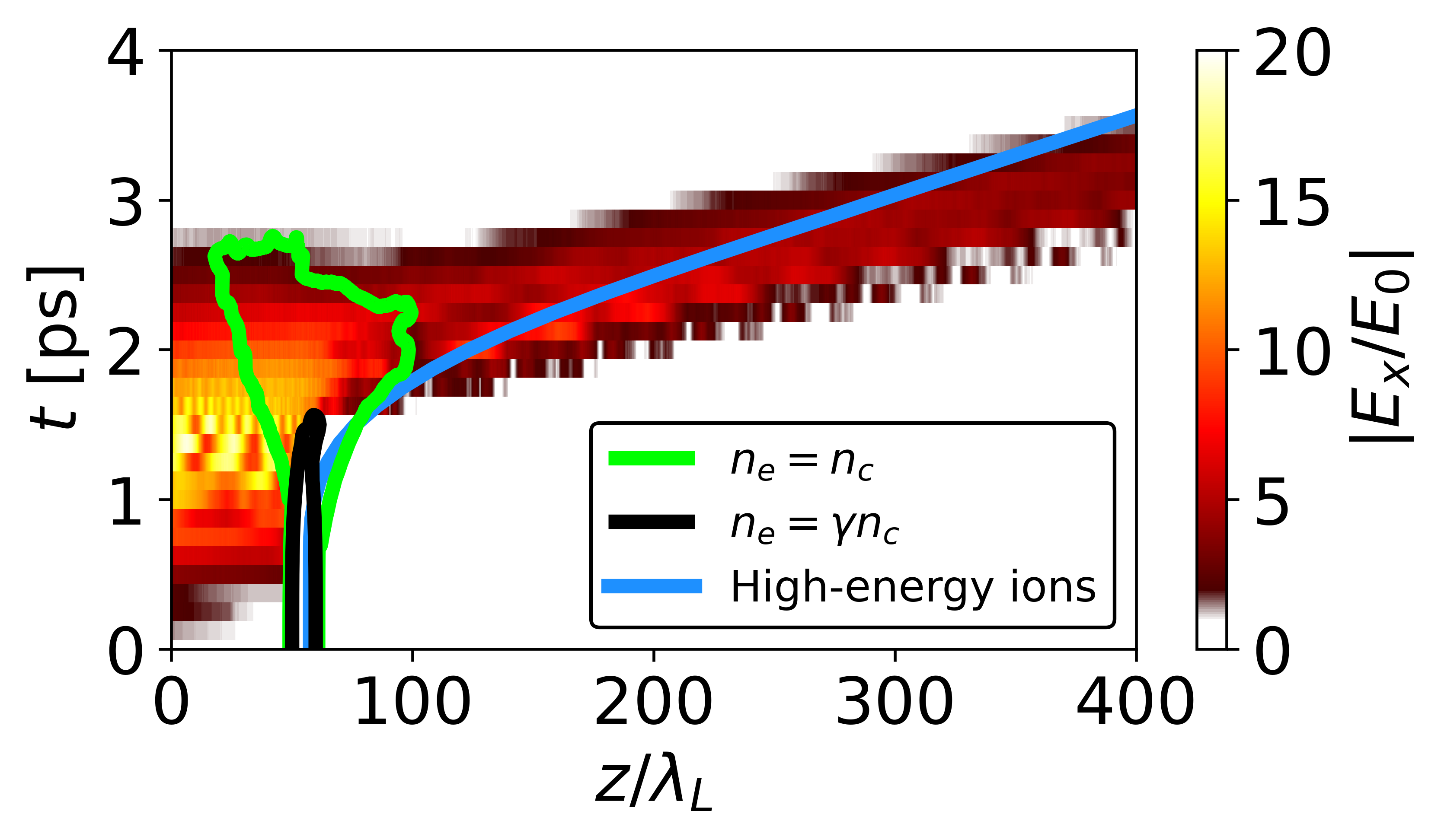}
    \caption{Space-time evolution of the on-axis laser amplitude ($E_{x}$) and electron density. Lineouts are averaged transversely and smoothed with a second-order Savitzky–Golay filter over $4\lambda_{L}$. Green and black contours mark $n_e=n_c$ and $n_e=\gamma n_c$. The average position of the highest energy protons is shown in blue.
    }
    \label{phase_space}
\end{figure}

We note that $\Theta_{0}$ can be matched to the electron heating model that works best for a given laser-target interaction. ESH acceleration remains optimized at $T^{*}=e^{-1}$ if the scaling for electron heating is linear with $T$. 

\begin{figure}
    \includegraphics[scale=0.5]{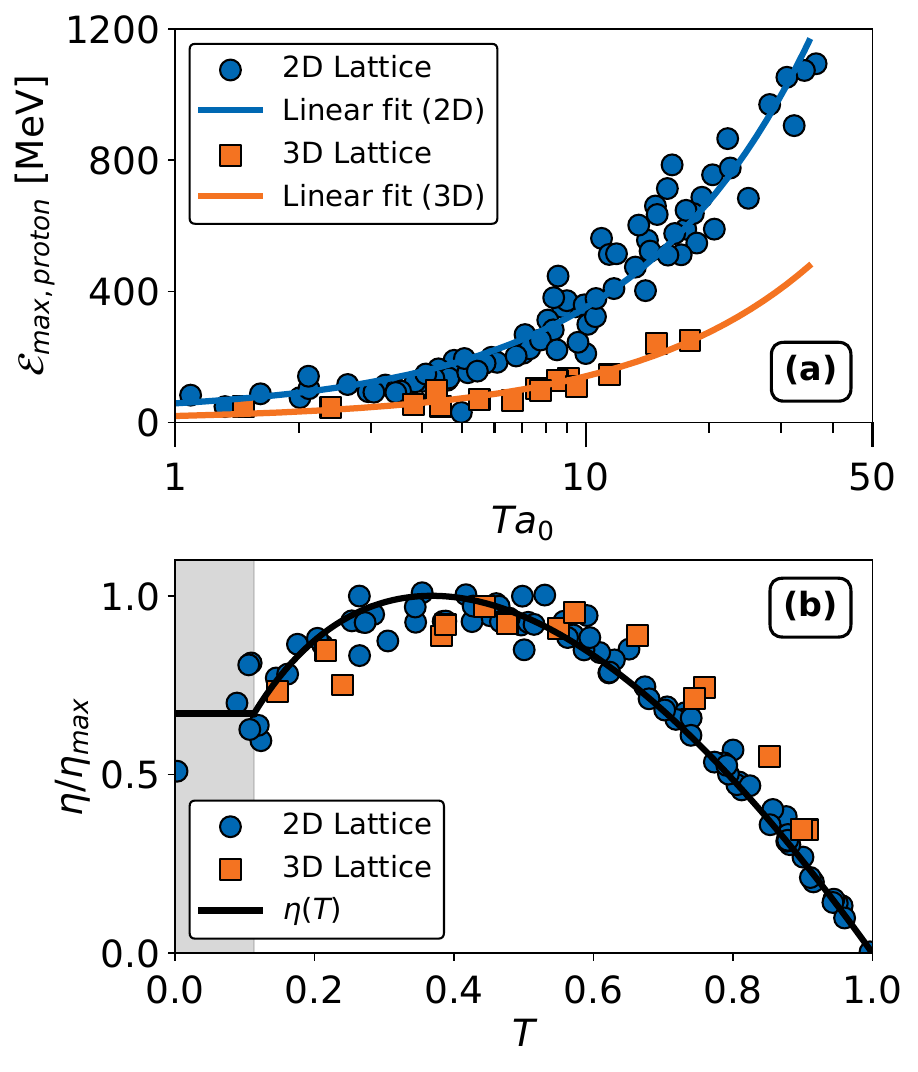}
    \caption{Results from 85 2D and 15 3D periodic-lattice simulations spanning $a_0=5$–$40$ and $\tau_p=0.5$–$2$~ps. Gray shading marks the TNSA regime $\Theta_e\le\Theta_{\mathrm{pond}}$. Panel (a) shows maximum proton energy versus transmitted field $Ta_0$ and panel (b) provides the combined proton-carbon efficiency versus $T$, normalized to $\eta_{\max}$ for each $(a_0,\tau_p)$.
    }
    \label{lattice-simulations}
\end{figure}

\begin{figure}[t]
    \centering
    \includegraphics[scale=0.5]{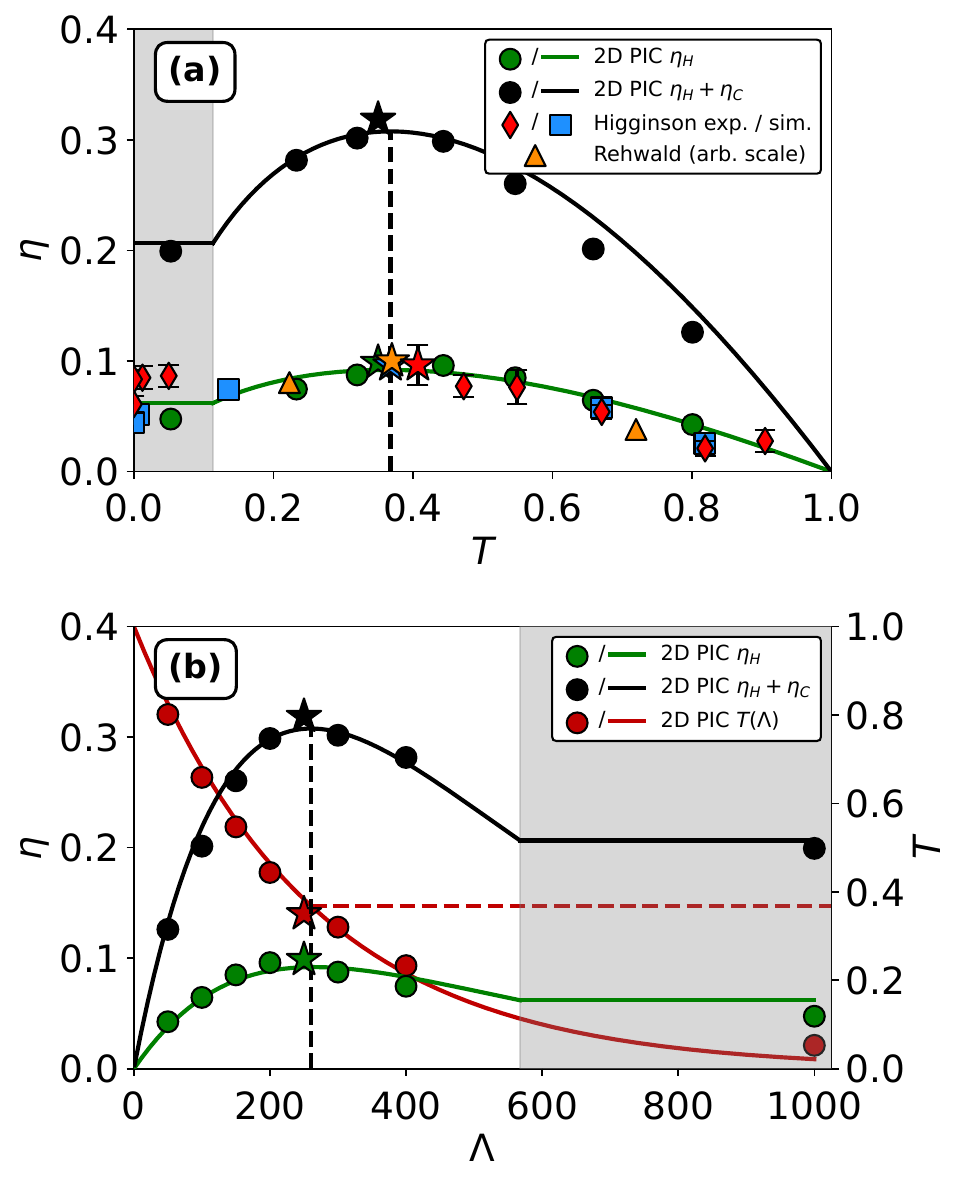}
    \caption{Ion conversion efficiency versus (a) $T$ and (b) $\Lambda$. Stars mark maximum efficiency for each data set and gray shading denotes the TNSA regime $\Theta_e\le\Theta_{\mathrm{pond}}$. Panel (a) shows experimental (red diamonds) and simulated (blue squares) data from Ref. \cite{higginson2018near} and arbitrary-scale data from Ref. \cite{rehwald2023ultra} (orange triangles; $T$ inferred from simulation). Simulation results from this work show proton (green) and proton-carbon (black) efficiency along with the attenuation model (red in panel (b)). The maximum at $\Lambda^*\approx250$ maps to $T^*=e^{-1}$.
    }
    \label{fig:2D-scaling-and-experiments}
\end{figure}

The proposed model for ESH is verified using the particle-in-cell (PIC) code SMILEI \cite{derouillat2018smilei}. This work leverages two types of PIC simulations: (a)~a full 2D model presented in Figs.~\ref{2D_PIC}, \ref{phase_space}, and~\ref{fig:2D-scaling-and-experiments} and (b)~2D and 3D periodic lattice models that allow large sweeps of plasma and laser parameters summarized in Fig.~\ref{lattice-simulations}. The full 2D simulations use a large domain of 150~\textmu m x 420~\textmu m with 50 cells/wavelength (taking $\lambda_{L}=1$~\textmu m). Periodic lattice simulations offer a significant reduction in computational resources at the cost of implementing a plane wave approximation. The periodic simulation grid is reduced to 10.26~\textmu m x 500~\textmu m in 2D and 3.0~µm x 3.0~µm x 500~µm in 3D with 50 cells/wavelength along the laser axis (i.e. $z$) and 50 (16) cells/wavelength along transverse directions (i.e. x, y) for 2D (3D).

Targets are initialized as a uniform, ionized CH$_2$ plasma matching a typical composition of hydrocarbon foams \cite{bailly-grandvaux2020ion,culfa2026proton}. Such conditions can be experimentally obtained by homogenizing aerogel or additively manufactured foams with an average density in the near-critical range \cite{tochitsky2025high,gu20253d}. Particle weights are selected so each macroparticle carries a density no more than $1n_c$ to $2n_c$ in 2D and $2.5n_c$ in 3D. Electrons (protons, carbons) are sampled up to 75 (25, 12) particles/cell for initial plasma densities spanning $0.1n_c$ to $100n_c$.

In the full 2D simulations the laser, injected at the left boundary, has a field strength $a_0=20$, a Gaussian spot size of $w_0=27.4$~\textmu m, and an intensity FWHM $\tau_p = 1$~ps for a total energy of 5.5 kJ (200 J/\textmu m in 2D). The laser electric field is linearly polarized along $x$ in all simulations. Here, a large spot size is selected to resemble the focus and pulse energy of a typical kJ-class laser like OMEGA-EP \cite{danson2015petawatt,iwata2021lateral}. An electron density of $n_e/n_c=25$ and length scale $L_{0}/\lambda_{L}=10$ are selected such that $\Lambda\approx\Lambda^{*}=260$ is optimized for ESH acceleration.

Figure~\ref{2D_PIC}(a)-(c) presents the full 2D optimized ESH mechanism at critical timesteps in the process. The colormaps show electron density (gray) and laser Poynting flux $\vec{S}=\vec{E}\times\vec{B}/\mu_0$ (orange) along with a sample of the highest energy protons (scatter points). Fields in all figures are normalized to an electric (magnetic) field $E_{0}$ ($B_{0}$) corresponding to the relativistic threshold $a_{0}=1$.

Ion acceleration begins with the expansion of an initially overdense plasma that establishes an electron sheath and launches hydrogen and carbon ions. Panel (a) at $t=1.5$~ps shows the plasma with peak intensity on-target where the highest energy protons are at the quasi-neutral boundary as typical in classical TNSA. In panel (b), relativistic effects allow the laser to break through the plasma where $n_e/n_c~<~\gamma=\sqrt{1+a_0^2/2}$. Continued longitudinal expansion and ponderomotive channeling drops the electron density along the laser axis to form a fully transparent channel and enable ESH ion acceleration. As evident in panels (b) and (c), protons accelerated by ESH within the sheath overtake the maximum energy obtained by ions at the quasi-neutral boundary. ESH ends when the laser decouples from the electron sheath.

Figure~\ref{2D_PIC}(d)-(f) provides the proton spectra, electron spectra, and proton phase space for both an optimized plasma ($\Lambda=250$) and a highly overdense plasma ($\Lambda=1000$) that is opaque to the laser and representative of classical TNSA. The phase space, sampled at $t=2.25$~ps, is paired with a lineout of the sheath field transversely averaged about the laser axis. ESH allows the optimized target to maintain an approximately isothermal electron population, in turn providing a stronger sheath field, enhanced ion energy, and increased ion flux. The sheath field is locally enhanced in regions of overlapping electron density and transmitted Poynting flux \cite{kim2022efficient}. Unlike TNSA, the highest energy protons do not coincide with the sheath edge where quasi-neutrality breaks down. Instead, high energy protons emerge within the sheath where $\Theta_{e}\partial_z\ln{n_e}$ is maximized. In contrast, when $\Lambda\gg\Lambda^{*}$ the classical TNSA sheath is depleted shortly after peak intensity arrives on-target. Electrons adiabatically cool as they exchange energy with the sheath and, consequently, TNSA ion acceleration plateaus as the sheath field rapidly decays in the expansion.

A full space-time picture of the ESH process for $\Lambda=250$ is presented by a streak diagram in Fig.~\ref{phase_space}. The cycle-averaged laser field and electron density are transversely averaged over $\pm2\lambda_{L}$ about the central axis of the laser. Similarly, the average location for a sample of 400 of the highest-energy protons was tracked during the simulation. The plasma is represented by contours for critical $n_e=n_c$ and relativistic-critical $n_e=\gamma n_c$ densities, clearly showing plasma expansion and the formation of a transparent channel. Once the channel forms, the attenuated laser field co-propagates with ions over an approximate length-scale $c\tau_{p}\gg~L_{0}$. The extended interaction length allows the laser to efficiently couple to ions by continuous and localized electron heating within the sheath.

A set of 85 2D and 15 3D periodic lattice simulations confirms the robustness of the ESH model by scanning a wide range of laser and plasma parameters. The scan covers nearly two decades of laser energy and intensity with $a_{0}~\in~[5,40]$ and $\tau_p~\in~[0.5, 2]$~ps. This covers relativistic intensities ranging from $3\cdot 10^{19}$ to $2\cdot 10^{21}$~$\text{W/cm}^{2}$ which are relevant to a number of modern kJ-class laser systems like OMEGA-EP, TITAN, PHELIX, and LFEX \cite{danson2015petawatt}. Scans of plasma parameters include a range of length scales $L_0/\lambda_{L}\in~[1,20]$ with an electron density range centered about the optimal areal density $\Lambda^{*}$.

Key metrics of ion acceleration performance, namely cutoff energy $\mathcal{E}_{\max}$ and conversion efficiency $\eta$, are well characterized by the transmission coefficient $T$ as shown in Fig.~\ref{lattice-simulations}. In panel (a), proton $\mathcal{E}_{\max}$ demonstrates a linear dependence on transmitted field strength $Ta_{0}$ which is consistent with the sheath field scaling $E_{z}\propto Ta_{0}$ \cite{mora2003plasma,mora2005thin}. Fig.~\ref{lattice-simulations}(b) verifies the scaling of $\eta$ as a function of $T$. $\eta$ represents the combined efficiency of proton and carbon acceleration and is normalized to $\eta_{\max}$ for each set of simulations of a given $\left(a_0,\tau_p\right)$.

To demonstrate experimental evidence of the ESH scaling, we turn to published experiments by Higginson et al. and Rehwald et al. that provide data for both ion acceleration efficiency and laser transmission \cite{higginson2018near,rehwald2023ultra}. Higginson et al. focus on thin foils and provide absolute measures of laser transmission through Al and CH targets. Fitting transmission data to an exponential attenuation law provides an estimate $C_{d}\Theta_{0}\approx~35.2$, within a factor of two of the 1D analytical estimate of $C_{d}\Theta_{0}=68^{+23}_{-27}$. The lower fitted value is reasonable given the tendency of the 1D treatment to overestimate electron heating and the substantial uncertainty in measured on-target intensity. Additionally, we use the experimental Rayleigh length $z_R=24$~\textmu m in place of $c\tau_p$ when computing $\Theta_{0}$ since $z_{R}\ll c\tau_{p}$ limits the maximum electron temperature. Proton conversion efficiency was measured for CH foils, allowing a direct mapping between $\eta$ and $T$ presented in Fig.~\ref{fig:2D-scaling-and-experiments}(a). Simulations from Higginson et al. also agree well with our PIC models despite vastly different length scales ($\approx100$~nm vs. 10~\textmu m). Fig.~\ref{fig:2D-scaling-and-experiments}(b) presents simulations from this work as a function of $\Lambda$ and demonstrates agreement with the attenuation law and the optimal areal density $\Lambda^{*}$. The scaling for $\eta$ is normalized to the PIC ion spectrum, which exhibits a flatter distribution compared to the self-similar Maxwellian in Eq.~\ref{ion_energy_spectrum}.

Similar measurements are reported by Rehwald et al. \cite{rehwald2023ultra}, which studies proton acceleration in cryogenic hydrogen jets. The use of a jet enabled high-repetition rate but the measurements include large statistical noise due to fluctuating overlap of the laser and target. Shots were down-selected for analysis based on measurement criteria for central hits, and three representative proton spectra are provided in the published data showing optimization about an intermediate jet diameter. Integrating the spectra provides a normalized proton conversion efficiency for different jet diameters. Jet diameter was measured by shadow size and is an effective surrogate for target areal density. Rehwald et al. offers simulation data to estimate the laser transmission as a function of diameter. Combining the data yields a normalized efficiency scaling as a function of $T$ that shows excellent agreement with the ESH model with a maximum near $T^{*}=e^{-1}$.

In conclusion, we presented an analytical model that predicts ion conversion efficiency based on laser and target parameters for laser-driven ion acceleration in relativistic-critical density plasmas. Our 1D model combines self-similar plasma expansion with partial laser transmission to promote electron heating, yielding a robust scaling over a wide range of laser intensities. We have confirmed our model with a large number of 2D and 3D PIC simulations for both planar wave fronts and finite laser spots. Experimental data from Higginson et al. \cite{higginson2018near} and Rehwald et al. \cite{rehwald2023ultra} agree with the conversion efficiency scaling. Together the experimental and simulation data provide strong evidence of the ESH mechanism, offering a clear path towards optimized and efficient ion acceleration in relativistic laser-plasma interactions.


This research is supported by DOE NNSA LRGF under the cooperative agreement DE-NA0003960. This work was performed under the auspices of the U.S. Department of Energy by Lawrence Livermore National Laboratory under contract DE-AC52-07NA27344. The authors would like to thank R. Simpson, E. Grace, D. Higginson, J. Vazquez, F. Fiuza, N. Lemos, S. Tochitsky, and R. Rajawat for their helpful feedback and discussions.

\bibliography{main}

\end{document}


\title{Supplemental Material for ``Efficient laser ion acceleration in near-critical density plasmas in the picosecond pulse regime''}

\author{J. Luoma}
\affiliation{School of Applied \& Engineering Physics, Cornell University}
\author{A. Kemp}
\affiliation{Lawrence Livermore National Laboratory}
\author{A. Longman}
\affiliation{Lawrence Livermore National Laboratory}
\author{D. Rusby}
\affiliation{Lawrence Livermore National Laboratory}
\author{G. Shvets}
\affiliation{School of Applied \& Engineering Physics, Cornell University}

\maketitle

Electron heating in intense, picosecond-scale laser pulses can generate super-ponderomotive temperatures. A robust scaling in the relativistic regime is provided by the stochastic heating model introduced by Miller et al. in Ref. \cite{miller2023maximizing}. Starting with a laser of normalized vector potential $a_0$, pulse duration $\tau_p$, and wavelength $\lambda_{L}$, the electron temperature is

\begin{equation}
    \Theta_{hot}=\left(\left(\frac{\varepsilon_D}{m_ec^2}\right)^{-1}+\left(\frac{L_{acc}\Theta_{pond}}{\lambda_{L}}\right)^{-1}\right)^{-1}.
    \label{Miller_scaling}
\end{equation}

The ponderomotive temperature for linear polarization is approximately $\Theta_{pond}\approx a_0/\sqrt{2}$ and $L_{acc}$ is the acceleration length scale. Constants $c$ and $m_e$ are the speed of light and electron mass. $\varepsilon_D$ is an energy diffusion scale derived in Appendix A of Ref. \cite{miller2023maximizing},

\begin{equation}
    \varepsilon_D=\frac{1}{2}a_0m_ec^2\sqrt{\frac{c\tau_p}{\pi\lambda_{L}}},
    \label{diffusion_energy}
\end{equation}

which sets an upper bound on electron temperature in the limit of long acceleration lengths or high intensity \cite{miller2023maximizing}. The temperature plateaus to the ponderomotive limit at low intensity. In our model, we simplify the expression for $\Theta_{hot}$ utilizing the fact that $L_{acc}\approx c\tau_p\gg\lambda_{L}$ holds for picosecond-scale laser pulses. This leads to a maximum normalized electron temperature,

\begin{equation}
    \Theta_0=\frac{a_0}{2}\sqrt{\frac{c\tau_p}{\pi\lambda_{L}}}.
    \label{theta_0}
\end{equation}

\begin{figure}[t]
    \centering
    \includegraphics[scale=0.6]{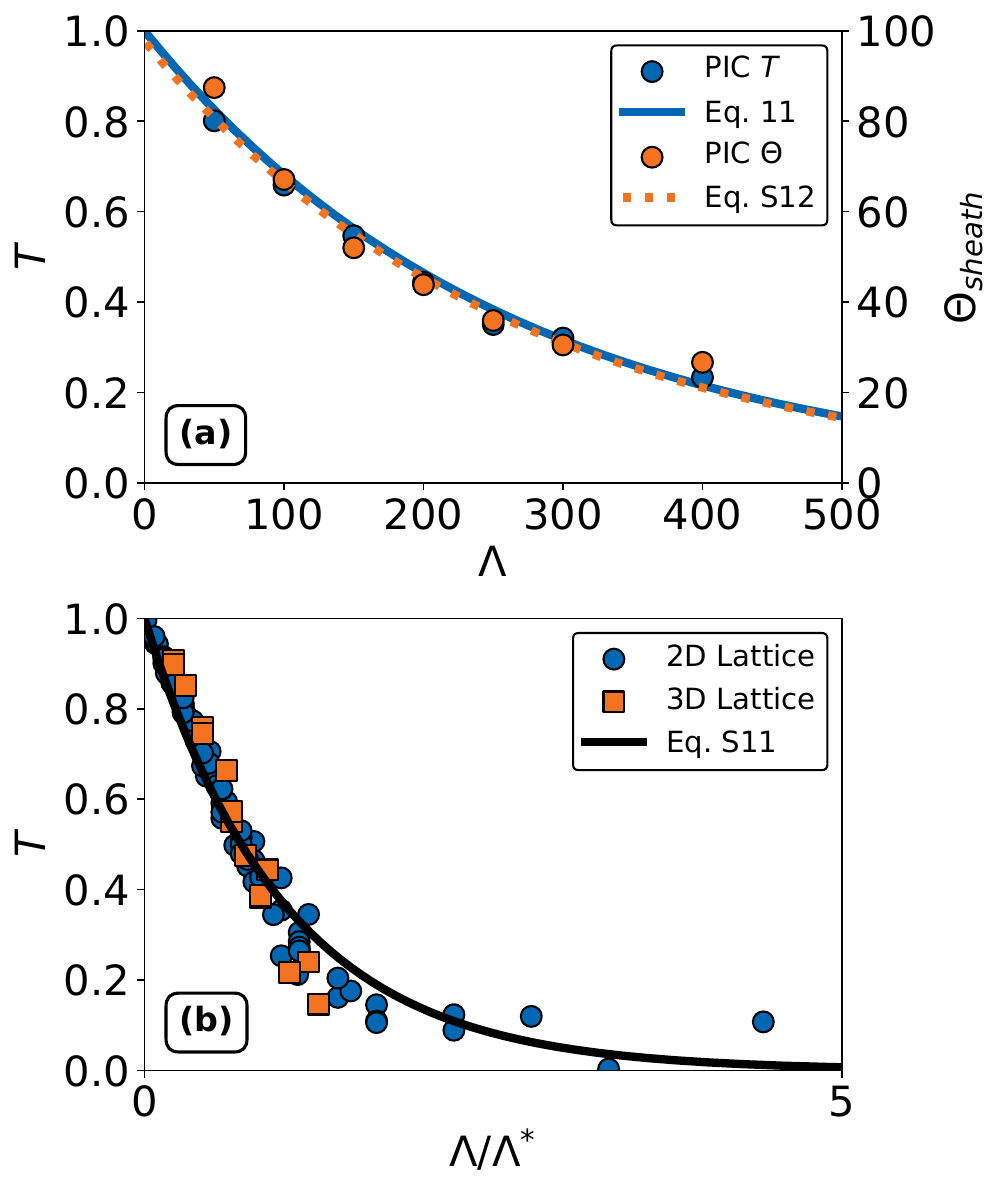}
    \caption{{Transmission and electron temperature scalings.} The (a) transmission coefficient and sheath electron temperature obtained in full 2D particle-in-cell (PIC) simulations show strong agreement with the analytical scalings for $T$ and $\Theta_e$, given by Eqs.~\ref{transmission_coeff} and \ref{sheath_temp}, respectively. The electron temperature is sampled at $t=2.25$~ps during the extended sheath heating process, which occurs after the laser channels through the target. The transmission scaling also agrees with (b) periodic-lattice 2D and 3D PIC simulations where $\Lambda^{*}=C_{d}\Theta_{0}$.}
    \label{fig:sheath_temp}
\end{figure}

Obtaining $\Theta_0$ requires that electrons interact with a laser field strength of $a_0$; however, energy absorption by the plasma will reduce the amplitude of the laser field and correspondingly lower the electron temperature. We account for this by considering a linear attenuation law with a constant attenuation coefficient $\alpha$ \cite{rybicki1979radiative},

\begin{equation}
    \frac{dI_{L}}{dz}=-\alpha I_{L}.
    \label{attenuation}
\end{equation}

Integrating intensity $I_{L}$ over the pulse envelope in time and transverse dimensions (i.e. x, y) allows the attenuation law to be expressed in terms of laser energy with $U_{L}=\int I_{L}dtdxdy$. An energy balance between $U_L$ and plasma energy~$U_{p}$ requires $\infrac{dU_{L}}{dz}+\infrac{dU_{p}}{dz}=0$. Hence we can express $\alpha$ as

\begin{equation}
    \alpha=-\frac{1}{I_{L}}\frac{dI_{L}}{dz}=-\frac{1}{U_{L}}\frac{dU_{L}}{dz}=\frac{1}{U_{L}}\frac{dU_{p}}{dz}.
    \label{attenuation_coeff_1}
\end{equation}

We consider the energy deposited into a thin slab of plasma with volume $dV=A_{p}dz$. The energy deposited into the plasma is the sum of electron energy $dU_{e}$ and ion energy $dU_{i}$, 

\begin{equation}
    dU_e=\frac{3}{2}\Theta_0m_ec^2n_{e0}A_{p}dz
    \label{electron_energy}
\end{equation}

and

\begin{equation}
    dU_{i}\approx\frac{1}{2}Mc_{s0}^2n_{i0}A_{p}dz=\frac{1}{2}\Theta_0m_ec^2n_{e0}A_{p}dz.
    \label{ion_energy}
\end{equation}

Electrons have a temperature $\Theta_{0}$ in a plasma of initial density $n_{e0}$ and area $A_{p}$. A factor of $3/2$ in the electron energy accounts for three degrees of freedom in our system \cite{iwata2025energy}. The differential plasma energy is given by

\begin{equation}
    \frac{dU_{p}}{dz}=2\Theta_{0}m_ec^2n_{e0}A_{p}.
    \label{plasma_energy}
\end{equation}

Over a differential step, the laser energy is near its nominal value. Modeling the laser with a temporal Gaussian envelope and substituting in $\Theta_{0}^2\propto a_0^2\tau_{p}$, the laser energy can be expressed as

\begin{equation}
    U_{L}=\frac{\pi^{3/2}}{\sqrt{\ln2}}\Theta_{0}^2m_ec^2n_{c}\lambda_{L}A_{L}.
    \label{laser_energy}
\end{equation}

The area $A_{L}$ accounts for integration over the spatial profile which is a Gaussian in full 2D simulations. The parameter $n_{c}=4\pi^2\epsilon_{0}m_{e}c^2/q_{e}^2\lambda_{L}^2$ is the critical plasma density where constants $\epsilon_{0}$ and $q_{e}$ are vacuum permittivity and electron charge. Solving for the attenuation coefficient $\alpha$ gives

\begin{equation}
    \alpha=\frac{2}{C_{d}\Theta_0}\frac{n_{e0}}{n_c\lambda_{L}}=\frac{2}{L_0}\frac{\Lambda}{C_{d}\Theta_0}.
    \label{attenuation_coeff_2}
\end{equation}

We introduce $\Lambda=n_{e0}L_{0}/n_{c}\lambda_{L}$ as the target normalized areal density for an initial length $L_0$. The constant $C_{d}$ accounts for integration over the laser spatiotemporal profile with $C_{2D}=\pi/\sqrt{2\ln{2}}$ used for full 2D simulations. Since periodic-lattice simulations inherently overestimate laser absorption, we fit simulation data to the attenuation model and find $C_{2D}^{PW}\approx3.7$ and $C_{3D}^{PW}\approx4.9$ for plane waves. Hence we find that laser attenuation can be expressed solely as a function of initial conditions. 

Sheath electrons located behind the target experience an attenuated field strength that is proportional to $\sqrt{I_{L}}$. Hence, we define the transmission coefficient $T=\sqrt{I_{L}/I_{L0}}$ where $I_{L0}$ is the initial laser intensity. The attenuation law in Eq.~\ref{attenuation} then yields

\begin{equation}
    T=\exp\left(-\frac{\alpha L_0}{2}\right)=\exp\left(-\frac{\Lambda}{C_{d}\Theta_{0}}\right).
    \label{transmission_coeff}
\end{equation}

We modify the electron temperature scaling of Eq.~\ref{theta_0} to account for the reduced laser field $Ta_0$ experienced by the sheath electrons. This results in a sheath temperature, 

\begin{equation}
    \Theta_{e}=\frac{Ta_{0}}{2}\sqrt{\frac{c\tau_{p}}{\pi\lambda_{L}}}.
    \label{sheath_temp}
\end{equation}

\begin{figure}[t]
    \centering
    \includegraphics[scale=0.58]{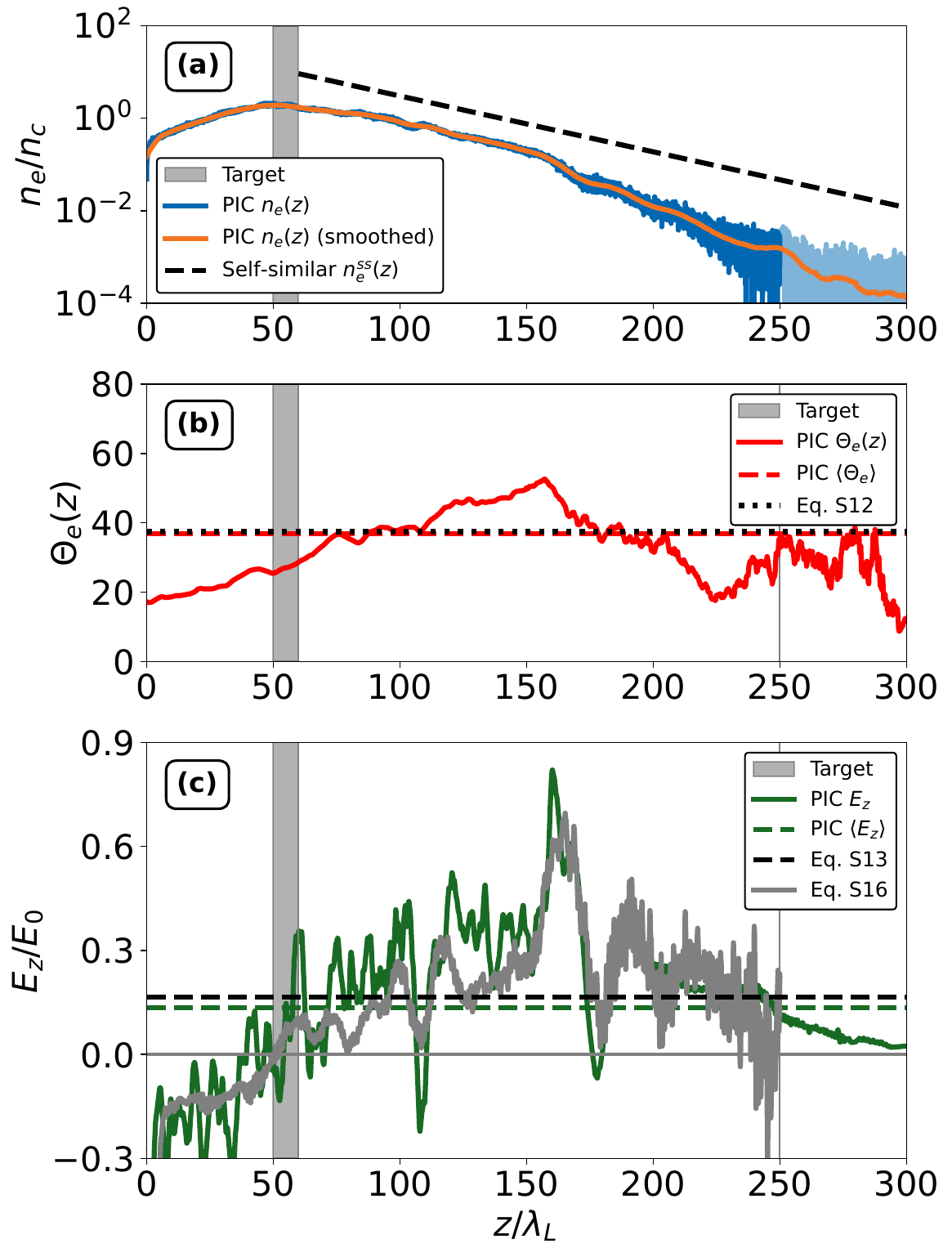}
    \caption{{Sheath field reconstruction.} The sheath field $E_z$ observed in full 2D PIC simulations is reconstructed using lineouts of (a)~electron density $n_e(z)$ and (b)~local temperature $\Theta_e(z\pm\Delta z)$ averaged over the laser spot size. Samples are taken during the extended sheath heating process at $t=2.25$~ps. A local temperature is obtained by averaging electron energy over a $\Delta z=\lambda_{L}$ window. The mean of $\Theta_e(z\pm\Delta z)$ over the sheath, spanning from the rear target surface at $z/\lambda_{L}=60$ to the quasi-neutral boundary at $z/\lambda_{L}\approx250$, shows excellent agreement with the isothermal approximation given by Eq.~\ref{sheath_temp}. Reconstructing the sheath, shown in panel (c), according to Eq.~\ref{sheath_eqn_z} demonstrates good agreement with PIC data.}
    \label{fig:sheath_reconstruction}
\end{figure}

The transmission law (Eq.~\ref{transmission_coeff}) and sheath heating model (Eq.~\ref{sheath_temp}) demonstrate excellent agreement with particle-in-cell (PIC) simulation data, as evident in Fig.~\ref{fig:sheath_temp}. The sheath temperature is obtained by sampling electrons in the sheath after the laser channels through the target at $t=2.25$~ps. Fitting the energy spectrum to a Maxwell-Boltzmann distribution yields the effective temperature of the sheath as presented in Fig.~\ref{fig:sheath_temp}(a). The exponential scaling of $T$ is robust across a wide range of laser and plasma parameters as shown in Fig.~\ref{fig:sheath_temp}(b), which includes simulation points covering $a_{0}~\in~[5,40]$, $\tau_p~\in~[0.5, 2]$~ps, and $L_0/\lambda_{L}\in~[1,20]$. The normalization $\Lambda/\Lambda^*$, where $\Lambda^*=C_d\Theta_0$ is the optimal areal density predicted by our model, collapses all of the simulation points to a single curve that characterizes transmission.

The remaining equations in our work utilize the self-similar solutions for plasma expansion as studied in Refs.~\cite{mora2003plasma} and \cite{gurevich1981ion}. The novelty of this work lies in coupling self-similar plasma expansion to a transmission-dependent electron temperature through the equation,

\begin{equation}
    |q_e|E_{z} =-|q_e|\partial_{z}\Phi = -\Theta_{e}m_{e}c^{2}\partial_{z}\ln{\left(n_e/n_{e0}\right)}.
    \label{sheath_eqn}
\end{equation}

This expression for $E_z$ assumes an isothermal sheath of temperature $\Theta_{e}$ and uses a self-similar solution for electron density given by $n_{e}^{ss}=n_{e0}\exp\left(-z/c_st-1\right)$. The transmission-dependent nature of $E_z$ provides a direct connection between the transmission coefficient $T$ and ion acceleration performance. Specifically, ion conversion efficiency can be expressed in terms of $T$,

\begin{equation}
    \eta = \frac{c_{s}t_{acc}}{L_{0}}T\ln{\left(T^{-1} \right)}.
    \label{ion_efficiency}
\end{equation}

The sound speed $c_s\propto\sqrt{T}$ might suggest that ion efficiency scales as $\eta\propto T^{\infrac{3}{2}}\ln{\left(T^{-1} \right)}$; however, the conservation of ion number requires $c_{s}t_{acc}\le L_{0}$ and sets an upper bound on efficiency. In the mass-limited regime $c_{s}t_{acc}=L_{0}$ and

\begin{equation}
    \eta=T\ln{\left(T^{-1} \right)}.
    \label{ion_efficiency_ml}
\end{equation}

The mass-limited scaling has a maximum at $T=T^{*}=e^{-1}$ with $\eta(T^*)\approx37\%$. This optimization condition and peak efficiency find good agreement with simulations of hydrocarbon plasmas, where the combined hydrogen and carbon ion efficiency reached $33\%$. Published experimental results also demonstrate good agreement with the optimization condition and proton conversion efficiency.

Relativistic transparency is a common feature of many enhanced ion acceleration schemes. For example, breakout afterburner (BOA) uses transparency to produce a two-stream instability between ions and fast electrons with modes that grow into an electrostatic accelerating field \cite{yin2011three,yin2011break,rajawat2016one,stark2018detailed}. Magnetic vortex acceleration (MVA) utilizes near-critical plasma channels to generate magnetic fields that produce strong accelerating fields upon expansion at the rear target surface \cite{nakamura2010high,tazes2024efficient}. In contrast, our approach does not require fields produced by instabilities nor magnetic field expansion. Instead we leverage the sheath field produced naturally from a simple plasma expansion that is dependent only on local electron temperature and density gradient. To illustrate this point, we reformulate the expression for the sheath field $E_z$ by relaxing the isothermal approximation for electrons by allowing $\Theta_e\xrightarrow{}\Theta_e(z\pm\Delta z)$ over a local scale $\Delta z$,

\begin{equation}
    E_{z}(z) = -\frac{m_{e}c^{2}}{|q_e|}\Theta_{e}(z\pm\Delta z)\partial_{z}\ln{\left(n_e(z)/n_{e0}\right)}.
    \label{sheath_eqn_z}
\end{equation}

Gradients in $\Theta_{e}$ add small contributions to $E_z(z)$ which are neglected. Profiles for $\Theta_e$ and $n_e$ are extracted from full 2D PIC simulations at $t=2.25$~ps by integrating over the transverse spot size of the laser. A localized electron temperature is obtained by assuming $\Theta_e(z\pm\Delta z)\approx\langle\gamma\rangle_{z\pm\Delta z}$ and averaging electron energy over $\Delta z=\lambda_{L}$. The density lineout is shown in Fig.~\ref{fig:sheath_reconstruction}(a) alongside the self-similar solution $n_{e}^{ss}(z)$. The density lineout is smoothed with a boxcar average with a $10\lambda_{L}$ window to obtain a differentiable profile. Since the PIC simulation allows 2D expansion of the plasma, the sheath density is expected to be lower than the 1D self-similar solution. Importantly, the density gradients of the PIC and self-similar solutions are in reasonable agreement. Fig.~\ref{fig:sheath_reconstruction}(b) provides lineout data for $\Theta_e(z)$ with dashed and dotted lines representing the (red) PIC average $\langle\Theta_e\rangle$ and (black) $\Theta_e$ as predicted by Eq.~\ref{sheath_temp}, which show strong agreement. Note we average PIC data from the target rear side at $z/\lambda_{L}=60$ to the quasi-neutral boundary at $z/\lambda_{L}\approx250$.

The sheath field is reconstructed according to Eq.~\ref{sheath_eqn_z} and presented in Fig.~\ref{fig:sheath_reconstruction}(c) alongside the PIC sheath field lineout. We truncate calculations at the quasi-neutral boundary at $z\approx250\lambda_{L}$. The reconstruction yields excellent agreement with the PIC lineout and demonstrates that localized electron heating directly enhances the sheath field. The mean PIC field $\langle E_z\rangle$ compares well to the self-similar solution Eq.~\ref{sheath_eqn}. This simple model clarifies the role of relativistic transparency, which is to locally increase $\Theta_e$ within the sheath and subsequently strengthen $E_z$. Decreasing target areal density $\Lambda$ allows for higher transmission and stronger sheath fields at the expense of reducing the number of ions available to accelerate. Hence, we balance these competing effects by optimizing for ion conversion efficiency with the scaling of Eq.~\ref{ion_efficiency_ml}.

\bibliography{main}